\documentclass[aps,pre,groupedaddress,showpacs]{revtex4}
\usepackage{graphicx}

\newcommand{\beq}{\begin{equation}}
\newcommand{\eeq}{\end{equation}}
\newcommand{\beqn}{\begin{eqnarray}}
\newcommand{\eeqn}{\end{eqnarray}}
\newcommand{\bmth}{\begin{displaymath}}
\newcommand{\emth}{\end{displaymath}}
\newcommand{\Rsc}{{\cal R}}
\newcommand{\om}{\omega}
\newcommand{\gb}{{\gamma_B}}
\newcommand{\lam}{\lambda}

\begin{document}

\title{$\cal R$-matrix theory of driven electromagnetic cavities}

\author{F. Beck}
\email{freder.beck@physik.tu-darmstadt.de}
\author{C. Dembowski}
\email{dembowski@ikp.tu-darmstadt.de}
\author{A. Heine}
\email{heine@ikp.tu-darmstadt.de}
\author{A. Richter}
\email{richter@ikp.tu-darmstadt.de}
\affiliation{Institut f{\"u}r Kernphysik, Technische Universit{\"a}t
Darmstadt, D-64289 Darmstadt, Germany}

\date{\today}

\begin{abstract}
Resonances of cylindrical symmetric microwave cavities
are analyzed in $\cal R$-matrix theory which transforms the input channel
conditions to the output channels. Single and interfering double resonances
are studied and compared with experimental results, obtained with
superconducting microwave cavities.
Because of the equivalence of the two-dimensional Helmholtz and the
stationary Schr{\"o}dinger equations, the results present insight into the
resonance structure of regular and chaotic quantum billiards.

\end{abstract}


\maketitle

\section{Introduction}

Nuclear reactions at low energies are characterized by narrow resonances which, in the case of well separated resonances, show a Breit-Wigner shape. If the separation is small, however, interference occurs which can lead to complicated non-homogeneous resonance shapes. This has been observed already in the early days of the study of slow neutron resonances. These data stimulated the theoretical development of resonance theories, as e.g. the stationary $\cal R$-matrix theory of Wigner and Eisenbud \cite{wig47}, or the equivalent one for running waves of Kapur and Peierls \cite{kap38}. The idea behind this work is a transformation from the input to the output channels without taking into account explicitely the complex dynamics of the strongly interacting system, which enters only through the parameters of the transformation matrix. These parameters have to be obtained by fitting the resulting multi-channel cross-section to experimental data. The essential benefit of the Wigner and Eisenbud formulation is the transformation to standing waves which results in a spatially stationary problem with real boundary conditions rather than the complex boundary conditions of running waves in each channel. Nevertheless, this procedure gets rather complicated when more than one resonance contributes significantly, and empirical analysis can be hindered considerably by insufficient statistics and large background contributions. A complete representation of the theoretical as well as the practical aspects of $\cal R$-matrix theory in nuclear physics which constituted the basis for the widespread later work in this field is the review article of Lane and Thomas \cite{lane58}. Although in a different context than in the present paper, $\cal R$-matrix theory has recently also been applied to multiphoton processes \cite{bur91,pur93}, in the analysis of Rydberg spectra \cite{hal93,ste96}, and to chaotic ionization of Rydberg states \cite{kru00,kru01}.

A sensitive test of single- and multi-level resonance theories can be obtained in the electromagnetic analogue of a quantum mechanical multi-channel system in the form of a flat superconducting cavity, fed with microwave radiation \cite{gra92,imapro}.
It has been shown \cite{gra92} that all resonances of such a system can be
resolved and analyzed with respect to their positions and widths. This allows a clean verification of level statistics for regular and chaotic cavities, resp., depending on the geometrical shape of the flat resonator. Superconducting cavities, with their high quality factors, can, however, also be analyzed with respect to line shapes of isolated and overlapping resonances which are almost undisturbed by additional broadening from wall losses leading to strongly overlapping resonances in the spectra, as in normal conducting resonators. This is typically due in the high frequency part of the spectra, relevant especially for chaotic systems \cite{imapro}. There, the 'true' resonances, i.e. not additionally broadened, still have widths smaller than the mean level distance which allows a theoretical approach on the basis of groups of isolated resonances. First analyses of resonances
with the Breit-Wigner formula have been performed in \cite{alt93,alt96}. We amend this work in two aspects. First, we start from the electromagnetic field conditions of microwave cavities, rather than assuming the Breit-Wigner formula as valid also there, and second, we use the $\cal R$-matrix formalism, rather than starting directly with the $S$-matrix, since the parameters of the $\cal R$-matrix allow direct modeling of resonance shapes and comparison of resonances in different channels.

In this paper, we first derive the resonance theory of a two-dimensional
microwave cavity, following the idea of Wigner and Eisenbud \cite{wig47}
to parametrize
the relation between input and output in the form of a multi-level reactance
matrix. From this, the $S$-matrix, giving the power transfer from input to
output, can be constructed by a non-linear transformation which leads, in
the multi-level case, to rather involved interference structures.

The approach is tested with experimentally determined single and double
resonances of a superconducting two-dimensional billiard of threefold
($C_{3}$) symmetry \cite{dreieck,dreieck2}. We also analyze single resonances
of a three-dimensional stadium billiard \cite{3dstad}. The tests are performed in the low as well as in the high frequency parts of the spectra, in order to cover, on one hand, strong resonances which extend over several orders of magnitude, and on the other hand the relevant frequency part for analyzing the dynamics of the system. At high frequencies level interference occurs, leading to characteristic non-uniform line shapes.
Comparing our one-level formula with the Breit-Wigner formula of nuclear physics shows that for narrow and isolated resonances the difference is neglibly small.

\section{Derivation of the cavity resonance formula}

\subsection{The cavity and antenna wave functions}

\begin{figure}
\centerline{\includegraphics[width=10cm]{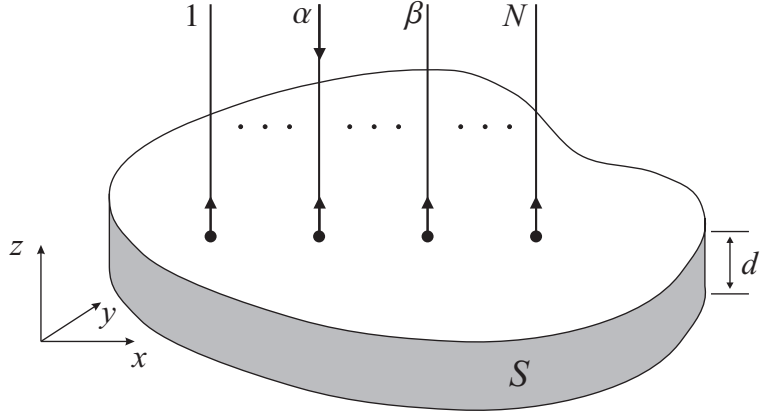}}
\caption{\label{abb1}
Sketch of the flat cavity with $N$ connecting antennas.
The feeding antenna is denoted by $\alpha$;
d: height of the cylindrical cavity, closed with two parallel plates
perpendicular to the cylinder; $S$: cylinder surface;
$z$ is the longitudinal,
and \{$x,y$\} are the transverse coordinates.
}
\end{figure}

We treat a cylindrical microwave cavity with $N$ antennas
$1,\dots \alpha, \dots \beta, \dots N$ connected perpendicular to it.
One antenna, which we always denote by $\alpha$, is feeding the cavity while radiation is extracted through the others (multi-channel situation, see Fig.~\ref{abb1}). The electromagnetic fields in the cavitiy can be derived from scalar fields \cite{jac75}, and it suffices to treat these. The complete scalar wave function with fixed frequency $\omega$ can be written as 
\beq
\label{fulwav}
\Xi (\vec{r},t)=\Psi(\vec{r}) \exp(-i\omega t)\,;\quad \vec{r}=\{x,y,z\}
\eeq
with the spatial part $\Psi(\vec{r})$ obeying the wave equation
\beq
\label{waveq}
(\Delta + \frac{\omega^2}{c^2})\Psi(\vec{r})=0\,.
\eeq 
For a cavity with arbitrary cross-sectional shape, but with end-plates perpendicular to the cylinder (cf. Fig.~\ref{abb1}), the spatial part $\Psi(\vec{r})$ can be separated into longitudinal and transverse components, according to 
$\Psi(\vec{r})=\psi(z)\chi(\vec{r}_t)$ with $z$, the longitudinal and $\vec{r}_t=x\cdot\vec{e}_1+y\cdot\vec{e}_2$, the transverse coordinates. The $z$-dependent part consists of standing waves, $A\,\sin kz+B\,\cos kz$, and the boundary conditions,
$\vec{E}_t=0$ at $z=0$ and $z=d$ for transverse magnetic (TM) fields lead with $k=p\,\pi/d$ to the form
\beq
\label{tmag}
E_z=\psi_{TM,p} (z)\,\chi (x,y)=\cos(\frac{p\pi z}{d})\,\chi (x,y)\, ,\quad p=0, 1, 2, \dots
\eeq
while for transverse electric (TE) fields the vanishing of $H_z$ at the $z$-boundaries requires
\beq
H_z=\psi_{TE,p} (z)\,\chi (x,y)= \sin(\frac{p\pi z}{d})\,\chi (x,y)\, ,\quad p=1, 2, 3, \dots\,.
\eeq
For flat cavities, i.e. $d$ much smaller than the wavelength of the
resonating microwaves in the cavity, TM modes are the only possible ones
since the $p=0$ modes alone can exist. 
This is the interesting case for a comparison with quantum mechanics
\cite{imapro}. However, the following derivation is more general and
holds for any value of $p$, thus describing arbitrary cavities with cylindrical geometry.
The transverse part $\chi (\vec{r})$ is a solution of the two-dimensional wave equation 
\beq
\label{traveq}
(\Delta_{t} + \frac{\omega_p^2}{c^2})\chi_p(\vec{r}_{t}) =  0\,;\quad \Delta_{t}\equiv\Delta - \frac{\partial\,^2}{\partial z^2}\,;\quad\omega_p^2=\omega^2- (\frac{c\, p\pi}{d})^2
\eeq
with the boundary condition on the cylindrical surface $S$
(cf. Fig.~\ref{abb1})
\beq
\label{boundc}
 \left.\chi_p\right |_S = 0 \, .
\eeq
Equation (\ref{traveq}) together with Eq.(\ref{boundc}), defines an eigenvalue problem for the two-dimensional transverse fields with eigenfunctions $\chi_{pr}(\vec{r_t})$ , $r=1,2,3,\dots$. For $p=0$, these fields are
 the direct analog of the stationary wave functions of a two-dimensional quantum billiard which makes the study of the flat electromagntic cavity interesting for illuminating the implications of quantum chaos. 
The TM fields are given by
\beqn
\vec{E}_t & = & -\frac{p\pi\, c^2}{d\,\omega_p^2}\,\sin(\frac{p\,\pi z}{d})\vec{\bigtriangledown}_t\,\chi_{pr} \nonumber \\
\vec{H}_t & = & \frac{i\,\omega\, c}{\omega_p^2}\,\cos(\frac{p\,\pi z}{d})\vec{\bigtriangledown}_t\,\chi_{pr} \nonumber .
\eeqn

The whole system can geometrically be divided into an {\em internal} and an {\em external} part where the internal part consists
of the cavity proper, while the external parts are the $N$ antenna wave guides. The concept of the $\cal R$-matrix approach
consits of obtaining a transformation from the feeding antenna to the output antennas by using the boundary conditions at the
connections between internal and external regions.

The longitudinal wave in an arbitrary antenna $\mu$ out of the $N$ antennas
in general consists of incoming and outgoing parts
\bmth
\psi_\mu (z_\mu) = A_\mu\, \exp(-ik_\mu z_\mu)+A_\mu'\, \exp(ik_\mu z_\mu)\,.
\emth
The transverse part $\chi_\mu(\vec{r}_{\mu t})$ obeys also the reduced wave equation
\beq
(\Delta_{\mu t} + \frac{\omega_\mu^2}{c^2})\chi_\mu(\vec{r}_{\mu t})=0\,;\quad \Delta_{\mu t}\equiv\Delta - \frac{\partial\,^2}{\partial z_\mu^2}\,;\quad\omega_\mu^2=\omega^2- c^2\,k_\mu^2,.
\eeq
For the group velocity of the running wave in antenna $\mu$ one obtains
\beq
v_{g\mu}=\frac{d\,\omega}{d\,k_\mu} = \frac{c^2k_\mu}{\omega}\,.
\eeq

We now specify to the case of $N$ open antennas where the input into antenna $\alpha$ produces outgoing fluxes in $\alpha$ (direct reflection) as well as in all other antennas $\beta\neq\alpha$. For this purpose we introduce local antenna coordinates $\vec{r}_\mu$ for an arbitrary antenna $\mu$ according to
\bmth
x_\mu=x-x_{\mu\, 0}\, ,\quad y_\mu=y-y_{\mu\, 0}\, ,\quad z_\mu=z-z_{\mu\, 0}
\emth
and $\{x_{\mu 0}, y_{\mu 0}, z_{\mu 0}\}$ denotes the footpoint of antenna $\mu$. For the geometry of Fig.~\ref{abb1} we have 
$z_{\mu 0}=z_0=d$ for all $N$ antennas.  
The spatial wave function, $\Psi(\vec{r})$, depends on which of the $N$ antennas has been chosen as the input
antenna, and thus contains incoming waves. For fixed frequency $\omega$ there are $N$ such distinct wave functions which
form a basis for the scalar cavity fields. To distinguish them they have to be marked by the index of the feeding antenna. With this notation the basis set consists of the $N$ functions $\Psi_\mu(\vec{r})\,;\, \mu = 1 \cdots N$. With our convention, which alwas denotes the input channel with $\alpha$, the spatial wave $\Psi_\alpha(\vec{r})$ approaches in antenna $\beta$ of the external region the form   
\beq
\Psi_\alpha(\vec{r}\rightarrow\vec{r}_\beta) = \psi_{\alpha\beta}(z_\beta)\chi_\beta(\vec{r}_{\beta t}) \quad\quad {\rm in\: antenna\: \beta}\,.
\eeq

Introducing the unitary $S$-matrix which transforms input into output we can write the longitudinal parts in the form
\beqn
\label{gl10}
\psi_{\alpha\beta}(z_\beta)& = & -S_{\alpha\beta}\,\exp(ik_\beta z_\beta)\,K_\beta\,;\quad \beta \neq \alpha \nonumber\\
\psi_{\alpha\alpha}(z_\alpha)& = & [\exp(-ik_\alpha z_\alpha)-S_{\alpha\alpha}\,\exp(ik_\alpha z_\alpha)]\,K_\alpha\nonumber\\
	& = &(\psi_{in}+\psi_{out})\,K_\alpha
\eeqn
where $K_\beta,\, K_\alpha$ are normalizing constants. For later purposes it is convenient to choose normalizations such that the ingoing flux per unit time is normalized to one, and the outgoing fluxes per unit time are given by the absolute squares of the $S$-matrix elements. 
This leads to
\beqn
P_\alpha\! ^{in}& = & K_\alpha^2\,\psi_{in}^\ast\psi_{in}\cdot\int\chi_\alpha^\ast\chi_\alpha d\sigma_\alpha\cdot v_{g\alpha}\stackrel{!}{=} 1\,;\quad
\int\limits_\alpha\chi_\alpha^\ast\chi_\alpha d\sigma_\alpha\equiv f_\alpha \nonumber\\
K_\alpha & = & \frac{1}{(f_\alpha v_{g\alpha})^{1/2}}=\frac{1}{c}(\frac{\omega}{f_\alpha k_\alpha})^{1/2}
\eeqn
\beqn
\label{ptrans}
P_{\alpha\beta}\!^{out}& = & K_\beta^2 |S_{\alpha\beta}|^2 f_\beta v_{g\beta}\stackrel{!}{=} |S_{\alpha\beta}|^2\,;\quad 
K_\beta = \frac{1}{(f_\beta v_{g\beta})^{1/2}}=\frac{1}{c}(\frac{\omega}{f_\beta k_\beta})^{1/2}\nonumber\\
P_{\alpha\beta}\!^{out}& = & |S_{\alpha\beta}|^2\quad\rightarrow\quad P_{\alpha\beta}\equiv\frac{P_{\alpha\beta}\!^{out}}{P_\alpha\!^{in}}=|S_{\alpha\beta}|^2\,.
\eeqn
where $\alpha$ is the input antenna, $\beta \neq \alpha$ are the $N-1$ output antennas, and $d\sigma_{\alpha}$ is the transverse surface element in antenna $\alpha$.
\subsection{Transformation to standing waves}
\noindent Analog to the formalism of Wigner and Eisenbud \cite{wig47} we perform a basis transformation to standing spatial waves $\Phi_\alpha(\vec{r})$, where the index $\alpha$ again indicates the input antenna, and the whole set $\{\Phi_\beta\,,\, \beta = 1 \cdots N \}$ is also complete for fixed $\omega$. This has the benefit to render the boundary conditions real. Because of the completeness of both sets we can expand the function $\Phi_\alpha$ into the set $\{\Psi_\beta\}$
\beq
\label{exp}
\Phi_\alpha=\sum_\beta C_{\alpha\beta}\Psi_\beta\,.
\eeq 

\begin{figure}
\centerline{\includegraphics[width=10cm]{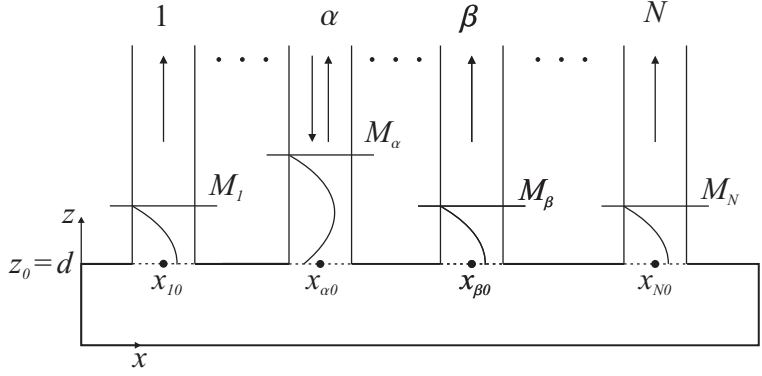}}
\caption{\label{abb2}
Perpendicular cut through the cavity at the antenna entrances. To
visualize the standing wave situation, virtual mirrors $M_i$ are placed
in the output antennas one quarter wavelength away from the antenna
footpoints in order to produce the boundary conditions as expressed
in Eq.(\ref{derant}). In the input antenna the mirror $M_\alpha$ has to
be placed at such a point that the frequency of the system stays at the
continously varying value $\omega$. This reflects itself in the
presence of sine and cosine waves in the entrance channel, Eq.~(\ref{stand}).
}
\end{figure}

The new base function $\Phi_\alpha(\vec{r})$ assumes, in analogy to Eq.(9), in the external region
the form 
\beq
\label{gl14}
\Phi_\alpha(\vec{r}\rightarrow\vec{r}_\beta)=\phi_{\alpha\beta}(z_\beta)\,\chi_\beta(\vec{r}_{\beta t})\quad {\rm in\: antenna\: \beta}
\eeq
with
\beqn
\label{stand}
\phi_{\alpha\beta}(z_\beta)& = & {\cal R}_{\alpha\beta}\,\cos (k_\beta z_\beta)Z_\beta\,\,\,\,\, \beta\neq\alpha \nonumber\\
\phi_{\alpha\alpha}(z_\alpha)& = & [k_\alpha^{-1}\sin (k_\alpha z_\alpha)+{\cal R}_{\alpha\alpha}\,\cos (k_\alpha z_\alpha)]Z_\alpha\,,
\eeqn
where $\cal R$ is the derivative (Wigner and Eisenbud \cite{wig47}), or reactance (as in most work on electromagnetic applications), matrix, and a factor $k_\alpha^{-1}$ has been put in for later convenience.
From Eq.(\ref{stand}) follow the derivatives at the antenna footpoints
\beq
\label{derant}
(\frac{d\phi_{\alpha\beta}}{d z_\beta})\mid_0=0\,;\quad (\frac{d\phi_{\alpha\alpha}}{d z_\alpha})\mid_0=Z_\alpha\,.
\eeq 
We can visualize this, see Fig.~\ref{abb2}, by placing virtual mirrors ($M_i$) into the antenna wave guides such that the system becomes stationary with standing waves in each antenna. The mirror is located in all exit antennas a quarter wavelength away from the antenna footpoint, while in the feeding antenna it has to be placed in such a way that the frequency of the system assumes the freely choosable value $\omega$.
 
Now, in the external region, we can identify the coefficients of the independent functions $\exp(ik_\mu z_\mu)$ and $\exp(-ik_\mu z_\mu)$ on both sides of Eq.(\ref{exp}) with the explicit forms of the wave functions in the external region, Eqs. (\ref{gl10}) and (\ref{gl14}), to arrive after a straightforward calculation at the connection between $\cal R$- and $S$-matrix which in matrix notation reads 
\beq
\label{smat}
S=\rho(1-iB{\cal R}B)^{-1}(1+iB{\cal R}B)\rho
\eeq
with the diagonal matrices
\beq
B_{\alpha\beta}=\delta_{\alpha\beta}k_\alpha^{1/2}\,;\quad \rho_{\alpha\beta} = \delta_{\alpha\beta} \exp(-i\,k_\alpha z_{\alpha 0})\,,
\eeq
and with
\beq
Z_\mu=k_\mu^{1/2} K_\mu=\frac{1}{c}(\frac{\omega}{f_\mu})^{1/2}\,.
\eeq
Equation (\ref{smat}) defines the nonlinear relation between the $S$-matrix and the reactance matrix $B{\cal R}B$.
\subsection{Derivation of the reactance matrix}
\noindent We now determine the reactance matrix from the relation between cavity eigenfunctions and antenna wave functions. This procedure is the essential part in the Wigner-Eisenbud approach, since it relates the $S$-matrix, Eq.(\ref{smat}), to the eigenstates of the cavity, without explicitly detailing its properties. We start with expanding the standing wave function $\Phi_\alpha$ inside the cavity into the complete orthonormalized basis of cavity eigenfunctions $\Psi_s$ which are the cavity eigenstates when all antennas are removed (closed internal region)
\beq
\label{expa}
\Phi_\alpha = \sum_s D_{\alpha s}\Psi_s \,,
\eeq
and from Eqs.(\ref{tmag}, \ref{traveq}, \ref{boundc}) the set $\Psi_s$ is defined (for transveres magnetic waves) by 
\beq
\label{cave}
\Psi_s (\vec{r})=\psi_{TM,p}(z)\,\chi_{pr}(x,y)\,;\: s \equiv \{p,r\}\,;\: p=0, 1, 2, \dots\,;\: r=1, 2, 3, \dots 
\eeq
where $p$ enumerates the longitudinal TM waves, and $r$ the transverse parts. The orthonormalization implies (though in our case of standing waves and time reversal symmetry all functions are real, we denote for generality the adjoint functions as in a complex function space)
\bmth
\int \Psi_{s'}^\star(\vec{r})\,\Psi_s (\vec{r})\,d^{3}r=\delta_{s's}\,,
\emth
 and the boundary conditions are (see Eqs.~(\ref{tmag}),(\ref{boundc})) 
\bmth
\chi_{rp} (x,y)\mid_S\,=\,0\,,\quad\frac{d\,\psi_{TM,p}}{d\,z}\mid_{z=0,d}\,=0\,.
\emth
The eigenfunctions $\Psi_s$ obey the eigenvalue equation
\beq
\label{waeq}
(\Delta + \frac{\omega_s^2}{c^2})\Psi_s = 0
\eeq
with eigenfrequencies $\omega_s$,
while the antenna wave functions have a continuous frequency spectrum
\beq
\label{antwaeq}
(\Delta + \frac{\omega^2}{c^2})\Phi_\alpha = 0\,.
\eeq
From Eqs.(\ref{expa}, \ref{waeq}, \ref{antwaeq}) we obtain
\beq
\label{detc}
\frac{1}{c^2}(\omega^2-\omega_s^2)D_{\alpha s}=\int\{\Phi_\alpha^\star \Delta \Psi_s - \Psi_s^\star \Delta \Phi_\alpha \}d^{3}r\,.
\eeq 
Applying Green's theorem
\bmth
\int(u \Delta v - v \Delta u)d^{3}r = \int (u\frac{\partial v}{\partial n} - v\frac{\partial u}{\partial n})d \sigma
\emth 
we can convert the right side of Eq.(\ref{detc}) into a surface integral which has contributions from the surface S and from the horizontal plates $z=0$ and $z=d$:
\beqn
\frac{1}{c^2} (\omega^2 - \omega_s^2)D_{\alpha s}=\int\limits_S [\Phi_\alpha^\star\frac{\partial\Psi_s}{\partial n}-\Psi_s^\star\frac{\partial\Phi_\alpha}{\partial n}]d\sigma \nonumber \\
+ \int\limits_{z=0}[\Phi_\alpha^\star\frac{\partial\Psi_s}{\partial n}-\Psi_s^\star\frac{\partial\Phi_\alpha}{\partial n}]d\sigma \nonumber \\
+ \int\limits_{z=d}[\Phi_\alpha^\star\frac{\partial\Psi_s}{\partial n}-\Psi_s^\star\frac{\partial\Phi_\alpha}{\partial n}]d\sigma\,
\eeqn
The first integral vanishes, because on $S$ both, $\Phi_\alpha$ and $\Psi_s$, are zero according to the boundary conditions.
The second integral also vanishes, since either ${\partial \Psi_s}/{\partial n}$ or ${\partial \Phi_\alpha}/{\partial n}$ are zero on the ($z=0$)-plane. The only non-vanishing contributions come from the third integral at each antenna footpoint($\mu = \alpha \dots N$). Because of the boundary conditions, Eq.(\ref{derant}), we obtain
\beqn
\frac{1}{c^2} (\omega^2 - \omega_s^2)D_{\alpha s}& = & -\sum_\mu \int\limits_\mu \Psi_s^\star \frac{d \phi_{\alpha\mu}}{d z_\mu}\mid_0 \chi_\mu(x_\mu,y_\mu) dxdy \nonumber \\
& = & - \frac{1}{c}(\frac{\omega}{f\alpha})^{1/2} f_{s\alpha} \psi_{TM,p}(d)\nonumber \\
f_{s\alpha}& = & \int\limits_\alpha \chi_s^\star(x_\alpha,y_\alpha)\cdot\chi_\alpha(x_\alpha,y_\alpha) dx_\alpha dy_\alpha\,,
\eeqn
since only the antenna $\mu=\alpha$ contributes to the sum.
From this we obtain for the expansion of Eq.(\ref{expa})
\beq
\label{expatwo}
\Phi_\alpha = c^2 \sum_s \frac{1}{\omega_s^2-\omega^2}\,\frac{1}{c}(\frac{\omega}{f_\alpha})^{1/2} f_{s\alpha}\cdot \Psi_s\,.
\eeq
Contracting each side of Eq.(\ref{expatwo}) for an arbitrary antenna $\beta$ with the transverse antenna wave $\chi_\beta^\star$ we obtain from Eq.(\ref{stand}), Eq.(\ref{expatwo}) and with $f_{s\beta}^\star=\int\limits_\beta \chi_\beta^\star(x_\beta,y_\beta)\cdot\chi_s(x_\beta,y_\beta) dx_\beta dy_\beta$
\beqn
\int\limits_\beta\chi_\beta^\star\Phi_{\alpha\beta}dx_\beta dy_\beta & = & {\cal R}_{\alpha\beta}\frac{1}{c}(\frac{\omega}{f_\beta})^{1/2}\int\limits_\beta \chi_\beta^\star\,\chi_\beta \,d x_\beta \,d y_\beta \nonumber \\
& = & {\cal R}_{\alpha\beta}\frac{1}{c}(\omega f_\beta)^{1/2} \nonumber \\                                                                                           
& = & c^2 \sum_s \frac{1}{\omega_s^2-\omega^2}\,\frac{1}{c}(\frac{\omega}{f_\alpha})^{1/2} f_{s\alpha}f_{s\beta}^\star\, [\psi_{TM,s}(d)]^2\,.
\eeqn
With
\beq
\gamma_{s\mu}=\frac{c}{f_\mu^{1/2}}f_{s\mu}\,\psi_{TM,s}(d)
\eeq
we finally get
\beq
\label{rmat}
{\cal R}_{\alpha\beta}=\sum_s \frac{\gamma_{s \alpha}\,\gamma_{s \beta}^\star}{\omega_s^2-\omega^2}\,,
\eeq
which is determined by the values of the cavity wave functions at the antenna footpoints.
If one uses the proportionality of energy and frequency in quantum mechanics ($E=\hbar\omega$) the difference of the reactance matrix $\cal R$ in the electromagnetic and in the quantum case \cite{wig47} is the resonance denominator. Here it contains the squares of the frequencies while in quantum mechanics the frequencies enter linearly.

The properties of the $S$-matrix for time reversal invariant systems, unitarity and reciprocity, which lead to relations between the $S$-matrix elements which are not linearly independent, enter in a simple way into the $\cal R$-matrix: $\cal R$ is a real and symmetric matrix which (for $N$ antennas) has $N(N+1)/2$ independent real parameters. Since for chaotic billiards one eventually wants to study systems which are not time reversal invariant, we have presented the derivation for the $\cal R$-matrix more generally, without using the reality condition.

\section{The experimental resonance spectrum}

As pointed out in Sec.~II. flat microwave cavities with cylindrical
symmetry are a powerful tool for experimental studies of two-dimensional
quantum billiards. While early experiments were carried out with
normal conducting devices at room temperature \cite{sri91,stoeckmann90},
only superconducting resonators allow measuring complete spectra in
a large frequency range with a high signal-to-noise
ratio \cite{gra92,imapro}.

\begin{figure}
\centerline{\includegraphics[width=10cm]{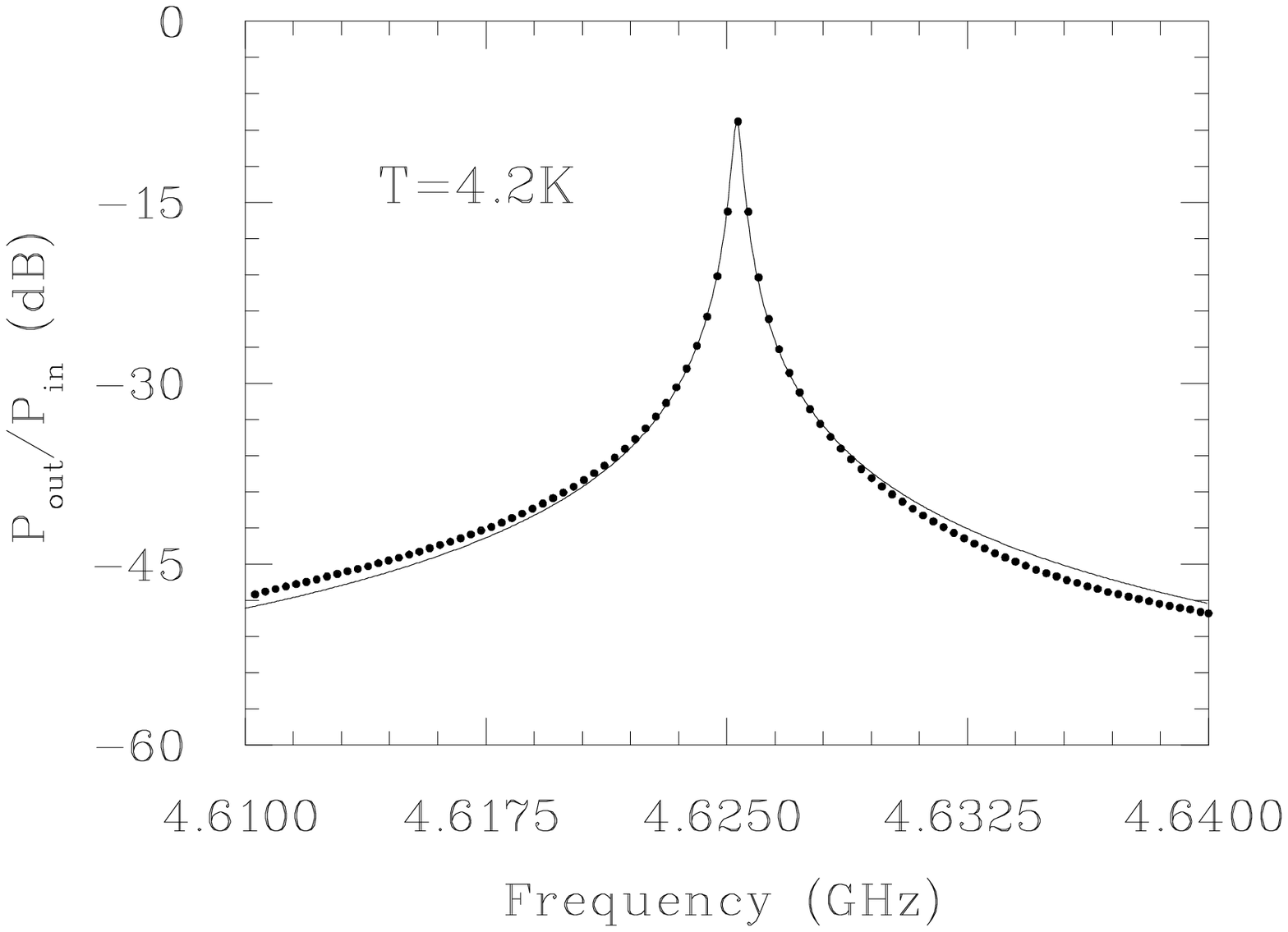}}
\centerline{\includegraphics[width=10cm]{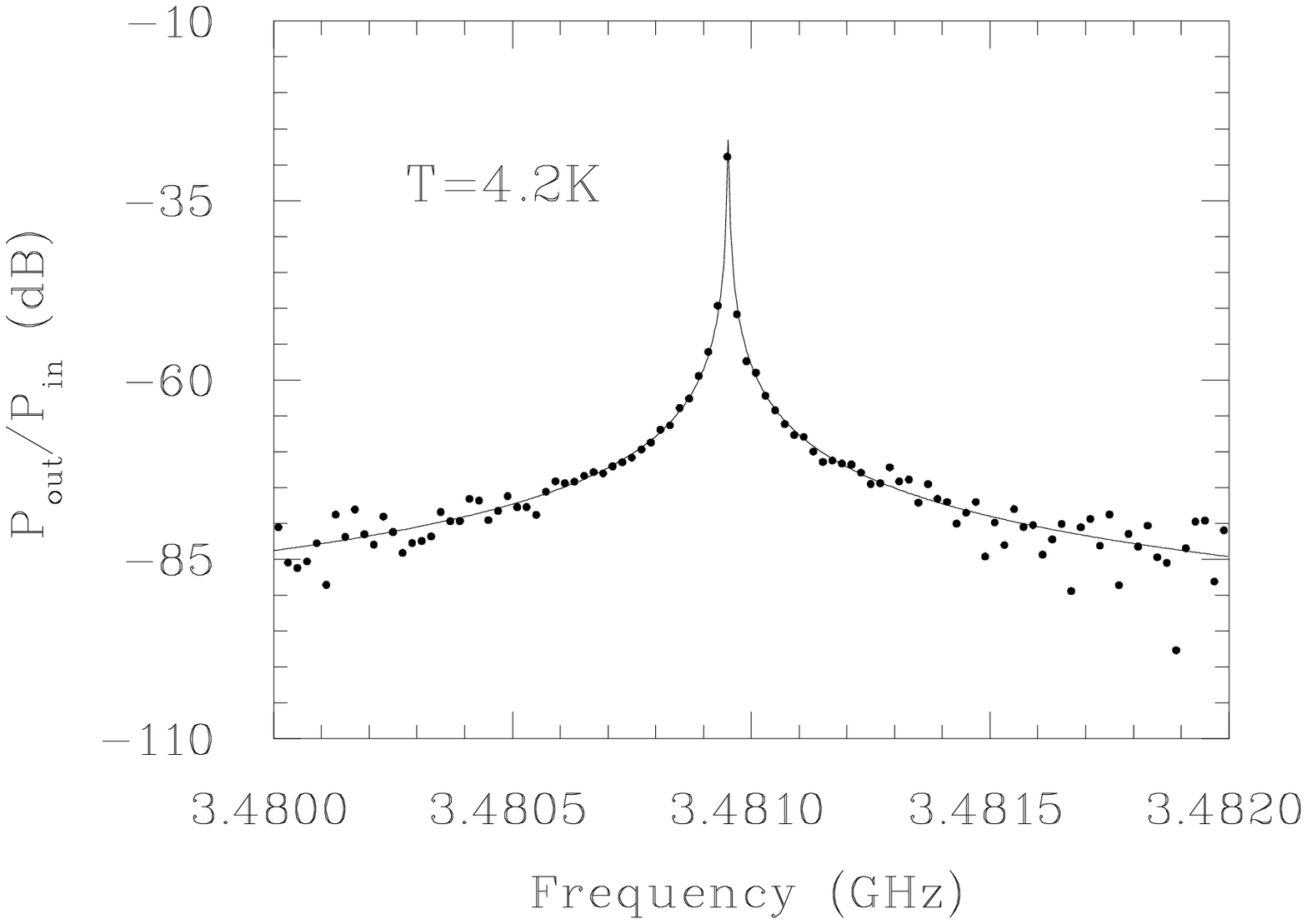}}
\caption{\label{abb3}
Comparison of two resonances measured in superconducting resonators 
with different quality factors.
Parameters of upper resonance (from \cite{dreieck}, 2D lead resonator):
$\Gamma_1 / 2\pi=91.7$ kHz,\
$\Gamma_2 / 2\pi=34.9$ kHz;\
$f_r=\omega_r / 2\pi=4.63$ GHz;\
$\lambda / 2\pi=84.1$ kHz;\
$\Gamma_{tot} / 2\pi=294.9$ kHz;\
$2\pi\tau=11.9\,\mu s$;\ fit accuracy $\chi^2=1.9$.\\
Parameters of lower resonance (from \cite{3dstad}, 3D niobium resonator):
$\Gamma_1 / 2\pi=0.02$ kHz,\ $\Gamma_2 / 2\pi=0.18$ kHz;\
$f_r=\omega_r / 2\pi=3.48$ GHz;\ $\lambda / 2\pi=0.51$ kHz;\
$\Gamma_{tot} / 2\pi=1.22$ kHz;\ $2\pi\tau=2.0$ ms;\
fit accuracy $\chi^2=9.7$.
}
\end{figure}

\begin{figure}
\centerline{\includegraphics[width=10cm]{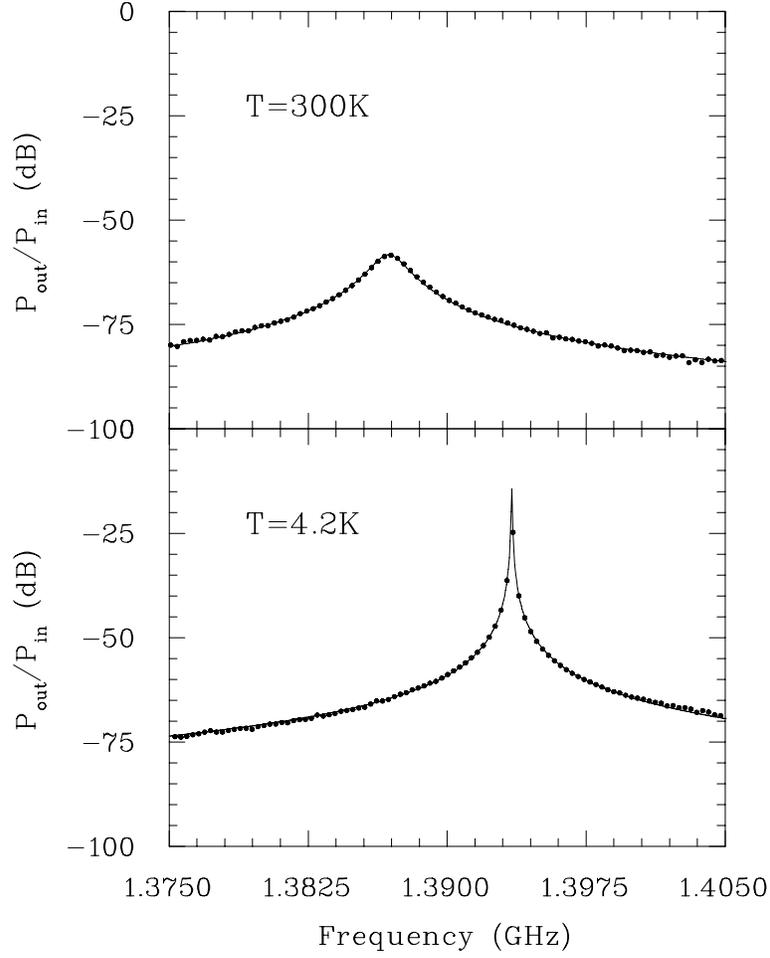}}
\caption{\label{abb4}
Comparison of a 'warm' ($T$ = 300~K) with a 'cold' ($T$ = 4.2~K) cavity,
showing the same isolated resonance in both cases. The shift of the
resonance frequency is due to the reduction of the size of the cavity
when cooling it to $T$ = 4.2~K. The fit curves coincide completely with the measured points. \\
Parameters of upper resonance:
$\Gamma_1 / 2\pi=1.3$ kHz,\
$\Gamma_2 / 2\pi=1.0$ kHz;\
$f_r=\omega_r / 2\pi=1.387$ GHz;\ 
$\lambda / 2\pi=949.2$ kHz;\
$\Gamma_{tot} / 2\pi=1900.6$ kHz;\
$2\pi\tau=1.1\,\mu s$;\ fit accuracy $\chi^2=1.07$.\\
Parameters of lower resonance (from \cite{dreieck}): 
$\Gamma_1 / 2\pi=3.8$ kHz,\
$\Gamma_2 / 2\pi=4.0$ kHz;\ 
$f_r=\omega_r / 2\pi=1.393$ GHz;\
$\lambda / 2\pi=10.3$ kHz;\
$\Gamma_{tot} / 2\pi=28.4$ kHz;\ $2\pi\tau=91.1\, \mu s$;\ 
fit accuracy $\chi^2=0.6$.
}
\end{figure}

\begin{figure}
\centerline{\includegraphics[width=10cm]{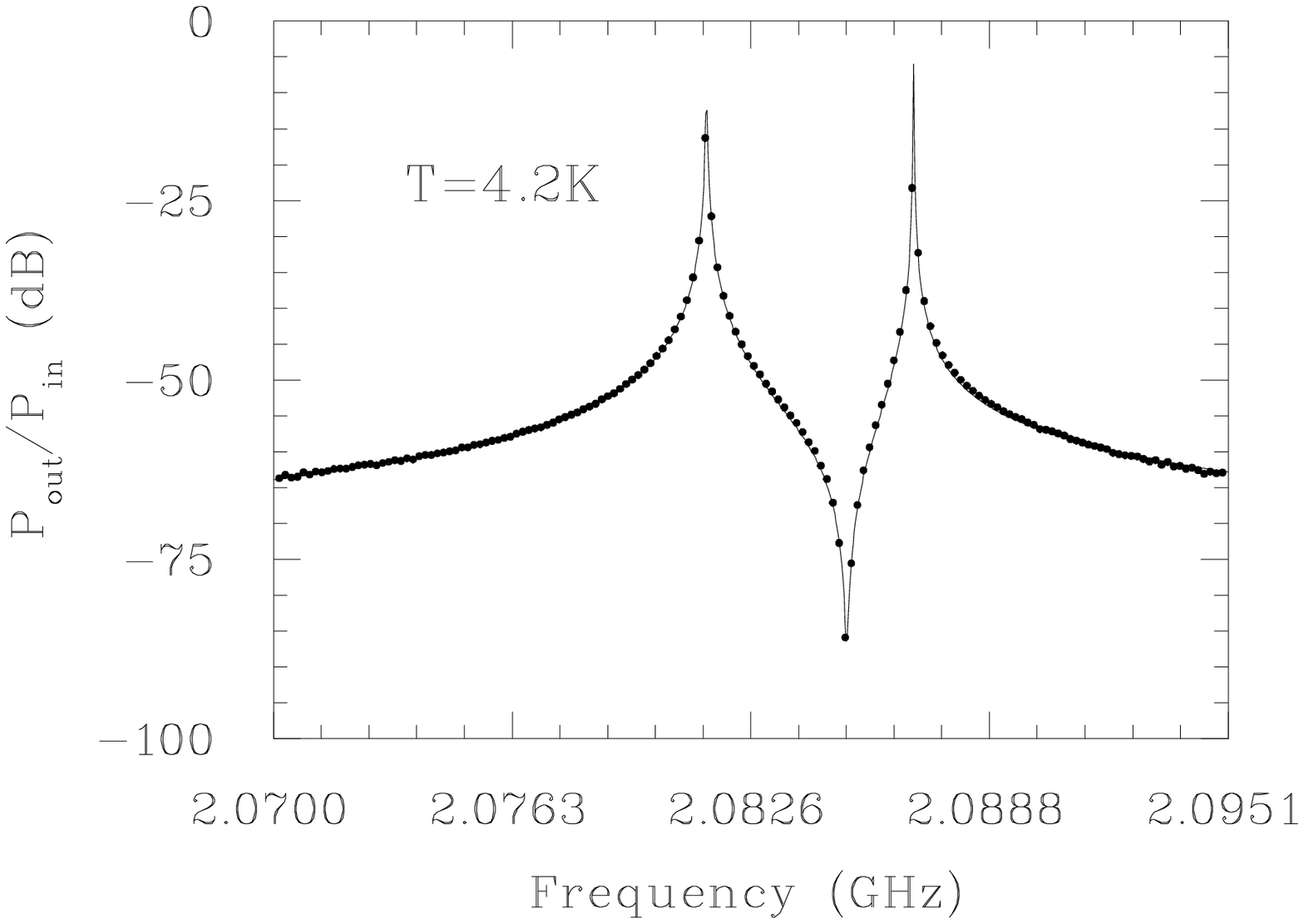}}
\centerline{\includegraphics[width=10cm]{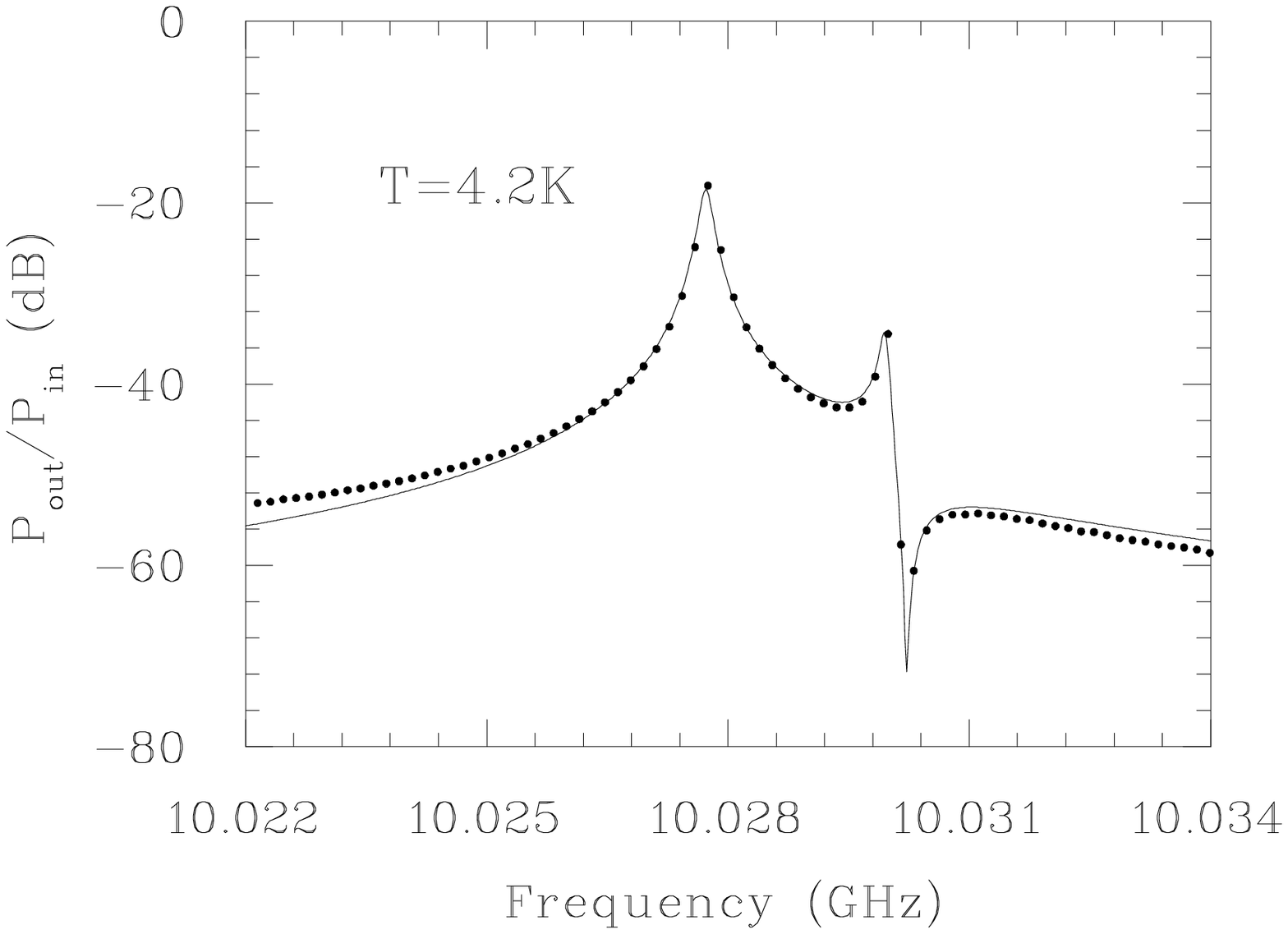}}
\caption{\label{abb5}
Two interfering double resonances (from \cite{dreieck}).\\
Parameters of upper resonance \\
first resonance:
$\Gamma_{11} / 2\pi=6.5$ kHz,\ 
$\Gamma_{12} / 2\pi=4.7$ kHz,\ 
$f_{1r}=\omega_{1r} / 2\pi=2.081$ GHz,\
$\lambda_1 / 2\pi=14.0$ kHz,\
$\Gamma_{1,tot} / 2\pi=39.1$ kHz;\
$2\pi\tau_1=71.4\, \mu s$ \\
second resonance:
$\Gamma_{21} / 2\pi=3.0$ kHz,\
$\Gamma_{22} / 2\pi=2.3$ kHz,\ 
$f_{2r}=\omega_{2r} / 2\pi=2.087$ GHz,\
$\lambda_2 / 2\pi=4.8$ kHz,\
$\Gamma_{2,tot} / 2\pi=14.9$ kHz;\
$2\pi\tau_2=209.9\, \mu s$; \\
fit accuracy $\chi^2=0.67$.\\
Parameters of lower resonance \\
first resonance:
$\Gamma_{11} / 2\pi=10.5$ kHz,\
$\Gamma_{12} / 2\pi=9.9$ kHz,\ 
$f_{1r}=\omega_{1r} / 2\pi=10.028$ GHz,\
$\lambda_1 / 2\pi=74.8$ kHz,\
$\Gamma_{1,tot} / 2\pi=170.0$ kHz;\
$2\pi\tau_1=13.4\, \mu s$ \\
second resonance:
$\Gamma_{21} / 2\pi=0.0033$ kHz,\
$\Gamma_{22} / 2\pi=34.7$ kHz,\
$f_{2r}=\omega_{2r} / 2\pi=10.030$ GHz,\
$\lambda_2 / 2\pi=29.7$ kHz,\
$\Gamma_{2,tot} / 2\pi=94.1$ kHz;\
$2\pi\tau_2=33.7\, \mu s$; \\
fit accuracy: $\chi^2=3.0$.
}
\end{figure}

\begin{figure}
\centerline{\includegraphics[width=10cm]{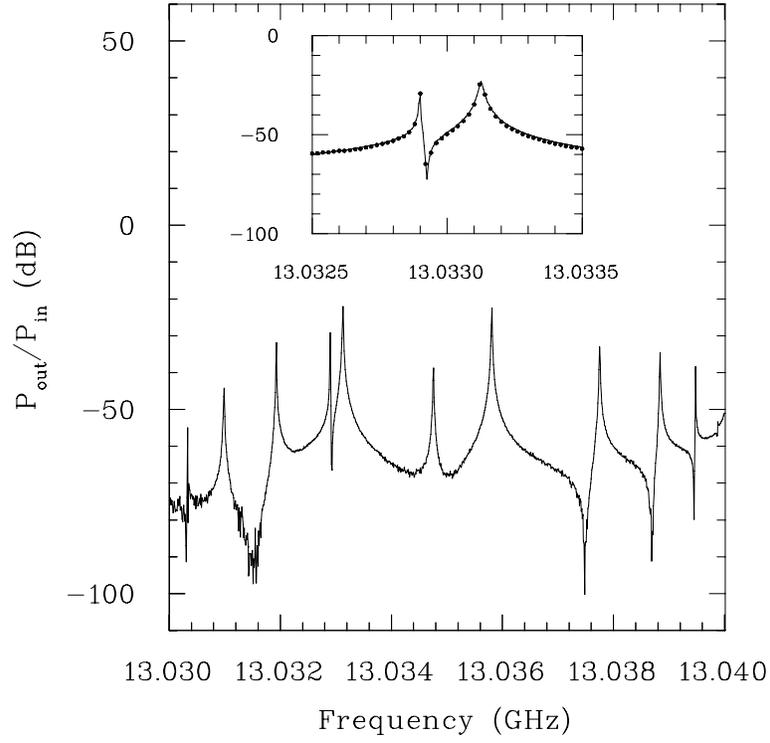}}
\caption{\label{abb6}
Part of the transmission spectrum of a 3D stadium billiard
\cite{3dstad} in the vicinity
of 13.03~GHz, where the resonances still can be described with the fit
formulae. Fit parameters for the double resonance shown in the insert,
first resonance:
$\Gamma_{11} / 2\pi=0.550$ kHz,\
$\Gamma_{12} / 2\pi=0.008$ kHz,\ 
$f_{1r}=\omega_{1r} / 2\pi=13.0329$ GHz,\
$\lambda_1 / 2\pi=1.44$ kHz,\
$\Gamma_{1,tot} / 2\pi=2.889$ kHz;\
second resonance:
$\Gamma_{21} / 2\pi=0.528$ kHz,\
$\Gamma_{22} / 2\pi=0.529$ kHz,\
$f_{2r}=\omega_{2r} / 2\pi=13.0331$ GHz,\
$\lambda_2 / 2\pi=7.16$ kHz,\
$\Gamma_{2,tot} / 2\pi=14.32$ kHz;\
fit accuracy: $\chi^2=1.1$.
}
\end{figure}

In \cite{dreieck} a chaotic billiard possessing threefold ($C_{3}$) symmetry
was studied experimentally. Its spectrum shows single as well as double
resonances (Fig.~\ref{abb3}, upper part, and Figs.~\ref{abb4} and \ref{abb5})
and thus provides a basis to test resonance formulae for
non-degenerate and nearly degenerate modes. The resonator used in the
experiment
is made from lead-plated copper and becomes superconducting below a
critical temperatur of $T_{c}$ = 7.2~K. The height of the cavity is
about 5.8~mm and therefore two-dimensional wave propagation is ensured
up to a critical frequency of about 25.5~GHz. The experiment was
carried out inside a liquid helium cryostat under stable conditions, where
at a temperature of 4.2~K the power transmitted from one antenna which
emits microwaves to a second antenna
which picks up microwaves, was measured for frequencies from 45~MHz to
25~GHz in steps of 10~kHz. The antennas consist of a wire that penetrates
the cavity through small holes in its lid by less than 0.5~mm.  
Due to the low rf resistance the superconducting
resonator possesses a quality factor $Q$ of the order of $10^{5}$ and
hence the lifted degeneracies of the double resonances can be resolved.
Even higher quality factors ($Q \approx 10^{7}$) can be achieved when
using electron-beam welded niobium resonators ($T_{c}$ = 9.2~K) or
three-dimensional microwave
billiards (see \cite{imapro}). The latter ones, however, do not show
an analogy between Helmholtz and Schr{\"o}dinger equation or
cylindrical symmetry,
but wave chaotic phenomena of the vectorial Helmholtz equation. Such a
billiard was studied in \cite{3dstad} and we analyzed here some of its
resonances for comparison (Fig.~\ref{abb3}, lower part, and Fig.~\ref{abb6}).

For our analyses we chose resonances in the low and high frequency parts of the spectra
(Figs.~\ref{abb3}, \ref{abb4}, \ref{abb5}, and \ref{abb6}).
We also measured one single resonance of the $C_{3}$
billiard at room temperature, i.e. with the normal conducting resonator.
This clearly shows a broader resonance line and a reduction
of the power transmitted by several orders of magnitude
(Fig.~\ref{abb4}).
A more detailed description of experiments with superconducting microwave
billiards can be found in \cite{imapro}. It should be emphasized, however,
that, though the measurement of a whole spectrum typically
takes one or more days, the data points in the vicinity of an individual
resonance are taken within a few seconds under identical circumstances.
This ensures the high precision of our experimental resonance lines.

\section{Analysis of single and double resonances}

For the analysis of experimentally determined resonances we augment the $\cal R$-matrix, Eq.(\ref{rmat}), in two aspects. First, we allow for the fact that the superconducting cavity is not a perfect conductor for electromagnetic waves. This results in an overall damping of the fields which are no longer strictly harmonic, as assumed in Eq.(\ref{fulwav}). We account for this by adding to the frequency a small imaginary part, $\omega_s\, \rightarrow\, \omega_s - i \lambda_s$, which results in an additional broadening of the resonances (cf. \cite{jac75}, Chap.8.8, where cavity losses because of finite wall conductivity are treated explicitely).

The introduction of a damping factor $\lambda$ produces a redundance problem for the extraction of the resonance parameters from the fitted spectra. Since the $\cal R$-matrix contains for two antennas and $s$ overlapping resonances
3$s$ independent parameters (the reduced widths, $\gamma_{s1,s2}$, and resonance frequencies, $\omega_{s}$), we introduce $s$ parameters more, the damping factors $\lambda_s$. This overdetermines the analysis since one can shift in the parameter space between the partial widths, $\Gamma_{si}=\gamma_{si}^{2}$, determining the resonance strengths and widths, and the damping factors $\lambda_s$, without changing the fit. We have carefully analyzed this problem analytically and numerically. The result is that the uncertainty introduced thereby does not influence the leading order of the partial widths and damping factors.

For the numerical analysis we set $c=1$ which then measures the wave numbers $k_\mu$ and the $\gamma_\mu$ in $s^{-1}$.

\subsection{Single resonances}

For single resonances the power transfer between one input and one output antenna, $P_{12}=\mid S_{12}\mid ^2$, is determined by Eq.(\ref{ptrans}) and Eq.(\ref{smat}) which connects the reactance matrix with the $S$-matrix. For a single resonance, viz. only one term in Eq.(\ref{rmat}), this leads to a resonance formula which resembles in its structure the Lorentz line (for the relation between this and the Breit-Wigner form, cf. Sec.~V)
\beq
\label{singr}
P_{12}(\omega)=\frac{4\,\gamma_1^2\, k_1\gamma_2^2\, k_2}{(\omega^2-\omega_r^2+\lambda^2)^2+(\gamma_1^2 \,k_1+\gamma_2^2\, k_2+2\lambda \omega_r)^2}
\eeq
where in the one level evaluation of Eq.(\ref{rmat}) the index $s=1$ has been obmitted. The partial widths, $\Gamma_i$, and the additional damping factor, $\epsilon$, are (the reason for this choice is presented in Sec.~V) 
\beq
\label{gam}
\Gamma_i=\gamma_i^2\,;\quad \epsilon=2\,\lambda
\eeq
which results in the total width
\beq
\label{gamt}
\Gamma_{tot}=\Gamma_1+\Gamma_2+\epsilon\,,
\eeq
and the lifetime of the resonance is given with the complex resonance frequency as $\tau=\lambda^{-1}$.

The evaluation of the parameters in Eq.(\ref{singr}) for a given cavity resonance is performed with the software package Mathematica 4.1, 'Nonlinear Fit'. For the antenna wave numbers $k_i$ we use the common value $k_i=\omega\,(i=1,2)$ which is the dominant mode for a coaxial cable \cite{jac75}.

Figure~\ref{abb3} shows two isolated cavity resonances (dotted  points) and their analysis with the theoretical relation, Eq.(\ref{singr}), (continuous line). The fit to the experimental points over a total interval of 4 to 7 orders of magnitude is nearly perfect, and this shows the reliability of the derived resonance formula. The resonance parameters are listed in the figure caption.

Figure~\ref{abb4} depicts the comparison of an isolated resonance in a
'warm', normally conducting cavity
with the same resonance in the superconducting cavity. The drastic increase
of the width is due to the additional wall damping. 
In the analysis this reflects itself in the vastly increased damping
factor, $\epsilon$. There is also a
shift of the resonance frequency, because the superconducting cavity
at T=4.2~K is contracted in size, and thus its
eigenfrequencies are higher (cf.~\cite{altetal}).
 For the resonance parameters, see the figure caption.

\subsection{Double resonances}

In this case the ${\cal R}$-matrix is given by (see Eq.~(\ref{rmat}))
\beq
\label{r2lev}
{\cal R}_{12}=\frac{\gamma_{11}\gamma_{12}}{\omega_1^2-\omega^2}+\frac{\gamma_{21}\gamma_{22}}{\omega_2^2-\omega^2}\,.
\eeq
The resulting expression for the $S$-matrix, Eq.(\ref{smat}), is extremely involved, containing a lot of nonlinaer interference terms, and it can not be simplified since no small parameter for an expansion can be defined. So, we  present the resulting formula for $P_{12}$ only in the Appendix. It is obtained by inserting Eq.(\ref{r2lev}) into Eq.(\ref{smat}), and this into Eq.(\ref{ptrans}).
Figure~\ref{abb5} shows two interfering double resonances and their analysis in ${\cal R}$-matrix theory. The two resonances exhibit very different interference stuctures which both are well described by the $\cal R$-matrix approach. The resonance parameters for both resonances are again listed in the figure caption, where the individual parameters for the two resonances have restricted significance because of the strong interference between both.

In Fig.~\ref{abb6} we present a section of the high frequency part of the spectrum of the 3D-cavity (from Ref.~\cite{3dstad}, more than 5000 resonances below this section!). It can be seen that the resonances are still well separated, leading to single and at most double resonances. The insert presents a fit to the double resonance around 13.033 GHz and shows the applicability of our analysis also in this part of the spectrum of a cavity which is not of a cylindrical
symmetric shape.

In all the analyses shown, the quoted fit accuracy $\chi^2$ refers to the
performed logarithmic fit of the logarithmic data.  

\section{Comparison of the cavity resonance with the Breit-Wigner formula}

For the comparison of the two 1-level formulas we start from
Eq.(\ref{singr}), i.e.
\bmth
P_{12}(\omega)=\frac{4\,\gamma_1^2\, k_1\gamma_2^2\, k_2}{(\omega^2-\omega_r^2+\lambda^2)^2+(\gamma_1^2 \,k_1+\gamma_2^2\, k_2+2\lambda \omega_r)^2}\;.
\emth
With $k_1=k_2=\omega$ and $\omega_{eff}^2:=\omega_r^2-\lambda^2$ one obtains
\beq
\label{sreff}
P_{12}(\omega)=\frac{4\,\gamma_1^2\,\gamma_2^2\,\omega^2}{(\omega-\omega_{eff})^2 (\omega+\omega_{eff})^2 + (2\,\lambda\, \omega_r+\gamma_1^2\omega+\gamma_2^2\omega)^2}\;.
\eeq

For narrow resonances $\Gamma, \lambda \ll \omega_r$ the response is non-vanishing only around $\omega\approx \omega_r$. Then we can set

\bmth
\omega=\omega_{eff}+\delta_1 \omega\,\quad \omega=\omega_r + \delta_2 \omega,\quad {\rm and}\quad \frac{\delta_1\omega}{\omega_r},\,\frac{\delta_2\omega}{\omega_r} \ll 1\;.
\emth

Up to first order in $\delta_1\omega/\omega$ and $\delta_2\omega/\omega$, Eq.~(\ref{sreff}), takes the form

\bmth
P_{12}(\omega)\approx \frac{\gamma_1^2\,\gamma_2^2}{(\omega-\omega_{eff})^2 (1-\frac{\delta_1\omega}{2\omega})^2+\frac{1}{4}[2\lambda(1-\frac{\delta_2\omega}{\omega})+\gamma_1^2+\gamma_2^2]^2}\,.
\emth

Neglecting the small corrections $\delta_1\omega/\omega,\,\delta_2\omega/\omega$, and with the partial and total widths of Eqs.(\ref{gam}, \ref{gamt}), $P_{12}(\omega)$ assumes the Breit-Wigner form

\beq
P_{12}(\omega)\approx \frac{\Gamma_1\,\Gamma_2}{(\omega-\omega_{eff})^2+ (\frac{1}{2}\Gamma_{tot})^2}\;.
\eeq

To give an example, for the first resonance of
Fig.~\ref{abb3} we have $\Gamma_{tot} / 2\pi=294.9$\ kHz,
$f_r=\omega_r / 2\pi=4.63$\ GHz and thus with
$\delta \omega\approx\Gamma_{tot}$,\ $\Gamma_{tot}/\omega_r=6.5\,\cdot 10^{-5}$. Consequently, the Breit-Wigner formula is   an excellent approximation for narrow single resonances in the electromagnetic case.

\section{Conclusion}

In this paper we study the resonance spectrum of a flat electromagnetic resonance cavity which is fed through an input antenna and analyzed by one or several output antennas. The TM eigenstates of the cavity obey the same Helmholtz eigenvalue equation as the two-dimensional stationary quantum problem in the same geometry (quantum billiard). Thus the study of cavity resonances provides, especially in the almost lossless superconducting case, an excellent tool to study experimentally the spectral properties of regular and chaotic quantum billiards. It is therefore of great importance for the analysis of the elctromagnetic case to derive a reliable resonance formula which not only reproduces the resonance shapes, but from which one can also deduce the relevant resonance parameters.

We derive the cavity resonance formula in close analogy to the Wigner-Eisenbud $\cal R$-matrix theory, which is a corner stone in nuclear reaction theory, and has been used in various fields, especially recently in atomic physics in the analysis of Rydberg atoms and multiphoton ionization. The goal of this approach is to express the $S$-matrix in terms of properties of the wave function at the channel entrancies which in our case are the antenna footpoints at the cavity. We succeed to express the cavity-antenna couplings in terms of the cavity eigenfunctions at the antenna entrances. As in the nuclear case, the resonance parameters can then be extracted by fitting measured resonance shapes. Differences to the nuclear case come from the different dispersion laws of the electromagnetic and quantum cases, and from the geometrical properties of TM resonances in the flat cylindrical cavity. For single, narrow resonances, however, the usual Breit-Wigner formula is an excellent approximation to the electromagnetic Lorentz lineshape.

We apply our multi-antenna result to measured well resolved single and 
double resonances of a superconducting cavity with one input and one output
antenna. The fits are almost perfect over many orders of magnitude of the
transmitted power. Rather involved interference patterns of double
resonances (cf. Fig.~\ref{abb5}) are well accounted for by the two-level
formula, even in the high frequency parts of the spectra.

\appendix*

\begin{acknowledgments}

We thank H.-D.~Gr{\"a}f for his ideas and suggestions concerning the
experiments. C.D., A.H., and A.R. thank H.~L.~Harney for valuable remarks,
concerning the numerical evaluation of the resonance formulas.
This work was supported by the DFG under contract no. Ri~242/16-3.

\end{acknowledgments}

\section{The fit formula for two antennas in the two-level case}

The power transfer $P_{12}$ from the input to the output antenna
is determined according to Eq.(\ref{ptrans}) by the absolute square
of the $S$-matrix element

\beq
\label{power}
P_{12}=\frac{P_{12}^{out}}{P_1^{in}}=\mid S_{12}\mid^2\;. 
\eeq

The $S$-Matrix is given according to Eq.(17) as

\beq
S=\rho(1-iB\Rsc B)^{-1}(1+iB\Rsc B)\rho\;.
\eeq

Since the diagonal phase factor $\rho$ does not contribute to the
absolute square we omit it in the following. Then we obtain with
$B\Rsc B \equiv M$

\beq
\label{sm}
S_{12}(\om)=\frac{2iM_{12}(\om)}{1-iM_{11}(\om)+
M_{12}(\om)M_{21}(\om)-iM_{22}(\om)-M_{11}(\om)M_{22}(\om)}\;.
\eeq

In order to simplify the evaluation we absorb the diagonal $B$-matrix
in the partial widths $\gamma_B$, and with $k_i=\omega$ (cf. Sec.\,IV)
we set

\beq
\gb_{,ij}(\om)=\sqrt{\omega}\, \gamma_{ij}\,;\quad (i,j)=(1,2)\;.
\eeq

Now we obtain from Eq.(34) for $B\Rsc B$

\beq
\label{reacmat}
M_{i,j}=\frac{\gb_{,1i}\gb_{,1j}}{\om_{1}^2-\om^2}+\frac{\gb_{,2i}\gb_{,2j}}{\om_{2}^2-\om^2}
\eeq

and, as pointed out in Sec.\,IV, a small imaginary part is added to the
frequencies

\bmth
\om_1=\om_{1r}-i\lambda_1\,;\quad \om_2=\om_{2r}-i\lambda_2\;.
\emth

Inserting Eq.(\ref{reacmat}) into Eq.(\ref{sm}) and taking the
absolute square $S^\star S$ one arrives at the extremely involved
result for $P_{12}$, Eq.(\ref{power}), for interfering double resonances.
For time reversal invariant systems which we analyze in this paper
exclusively, one obtains

\bmth
P_{12}(\omega)=\frac{N(\om)}{D(\om)}
\emth

with

\beqn
N(\om)=\lefteqn{4\left\{\left[2\gb_{,21}\gb_{,22}\lam_1\om_{1,r}+2\gb_{,11}\gb_{,12}\lam_2\om_{2,r}\right]^2+\left[\gb_{,21}\gb_{,22}(\lam_1^2+\om^2-\om_{1,r}^2)\right.\right.}\hspace{12cm}\nonumber\\\lefteqn{\left.\left.+\gb_{,11}\gb_{,12}(\lam_2^2+\om^2-\om_{2,r}^2)\right]^2\right\}^2\;,}\hspace{12cm}\nonumber
\eeqn
\beqn
D(\om)=\lefteqn{\left[\lam_2^2(\gb_{,11}^2+\gb_{,12}^2+2\lam_1\lam_2^2\om_{1,r})+(\gb_{,11}^2+\gb_{,12}^2)\om^2+2\lam_1\om^2\om_{1,r}\right.}\hspace{12cm}\nonumber\\\lefteqn{+(\gb_{,21}^2+\gb_{,22}^2)(\lam_1^2+\om^2-\om_{1,r}^2)+2\lam_2(\lam_1^2\om_{2,r}+\om^2\om_{2,r}-\om_{1,r}^2\om_{2,r})}\hspace{12cm}\nonumber\\\lefteqn{\left.-(\gb_{,11}^2+\gb_{,12}^2)\om_{2,r}^2-2\lam_1\om_{1,r}\om_{2,r}^2\right]^2}\hspace{12cm}\nonumber\\\lefteqn{\left[\gb_{,11}^2\gb_{,21}^2-2\gb_{,11}\gb_{,12}\gb_{,21}\gb_{,22}+\gb_{,11}^2\gb_{,22}^2-\lam_1^2\lam_2^2-(\lam_1^2+\lam_2^2)\om^2\right.}\hspace{12cm}\nonumber\\\lefteqn{-\om^4+2(\gb_{,21}^2+\gb_{,22}^2)\lam_1\om_{1,r}+\lam_2^2\om_{1,r}^2+\om^2\om_{1,r}^2+2\gb_{,11}^2\lam2\om_{2,r}}\hspace{12cm}\nonumber\\\lefteqn{\left.+2\gb_{,12}^2\lam_2\om_{2,r}+4\lam_1\lam_2\om_{1,r}\om_{2,r}+(\lam_1^2+\om^2-\om_{1,r}^2)\om_{2,r}^2\right]^2\;.}\hspace{12cm}\nonumber
\eeqn

Setting $\gb_{,21}=\gb_{,22}=\lam_2=\om_{2,r}=0$ and
dividing numerator and denominator by $\om^4$ one obtains the 
single resonance formula of Eq.(\ref{singr}).

In the parameter search of the fitting procedure one has to assure that
the decay parameters $\lam_1, \lam_2$ stay positive. This can be
achieved by setting $\lam_i=(\lam_i^2)^{(1/2)}$.

\newpage

\newpage

\newpage

\newpage

\newpage

\end{document}